\documentclass[prb,twocolumn,preprintnumbers,amsmath,amssymb,floatfix,reprint]{revtex4}
\usepackage{hyperref}
\usepackage{epsfig}
\usepackage[usenames,dvipsnames]{xcolor}
\begin{document}

\title{Optimal energy dissipation in sliding friction simulations}

\author{A. Benassi$^{1,2}$, A. Vanossi$^{3,4}$, G.E. Santoro$^{3,4,5}$ and E. Tosatti$^{3,4,5}$}

\affiliation{
$^1$ CNR Istituto per l'Energetica e le Interfasi (CNR-IENI), Via Cozzi 53, I-20125 Milano, Italy \\
$^2$ Centro S3, CNR Istituto Nanoscienze, Via Campi 213/A, I-41125 Modena, Italy \\
$^3$ International School for Advanced Studies (SISSA), Via Bonomea 265, 34136 Trieste, Italy \\
$^4$ CNR-IOM Democritos National Simulation Center, Via Bonomea 265, 34136 Trieste, Italy \\
$^5$ International Centre for Theoretical Physics (ICTP), P.O.Box 586, I-34014 Trieste, Italy
}

\date{\today}

\begin{abstract}
Non-equilibrium molecular dynamics simulations, of crucial importance in sliding friction, are hampered
by arbitrariness and uncertainties in the removal of %phonon
the frictionally generated Joule heat.
Building upon general pre-existing formulation,
we implement a fully microscopic dissipation approach which, based on a parameter-free, non-Markovian,
stochastic dynamics, absorbs Joule heat
equivalently to a semi-infinite solid
and harmonic substrate.
As a test case, we investigate the stick-slip %motion
friction
of a slider over a two-dimensional Lennard-Jones solid,
%and
comparing our virtually exact frictional results with approximate ones from commonly adopted
dissipation schemes. Remarkably, the exact results can be closely reproduced by a standard Langevin
dissipation scheme, once its parameters are determined according to a general and self-standing
variational procedure.
\end{abstract}

\maketitle

\section{Introduction}
Ordinary, macroscopic sliding friction, a far reaching subject  of enormous physical, technological and practical
importance, is notoriously complex and hard to approach from a microscopic viewpoint, both experimentally
and theoretically.  The two last decades have seen quiet but important progress in that arena. Experimentally,
the advent of nanosize slider methodologies is offering much fresh data and lively progress.  On the theory
side, advances in computing hardware and codes now allows atomistic molecular dynamics (MD) simulations
to be extensively used to describe sliding nanofriction: not simply as a mean of supplementing experimental
studies, but as a general framework for gaining unique insight into the relevant tribological processes
%for extended tribological systems,
sometimes overturning conventional wisdom~\cite{persson_book,robbin_muser}.
In MD simulations, the classical dynamics of atoms is described by solving numerically Newton's
equations of motion in a controlled computational experiment, where the interface geometry,
sliding, boundary conditions and inter-particle interactions can be chosen and varied to explore
various effects on friction, adhesion and wear. By following the particle dynamics for a significant
amount of time, quantities of physical interest such as instantaneous and average frictional force,
mean velocities, heat flow, and correlation functions are calculated to characterize the sliding motion
and the corresponding steady-state values. Unlike standard equilibrium MD simulations, %of bulk systems,
friction modeling inherently involves dynamics and properties quite far from equilibrium. Moreover,
as a rule, the dynamics is highly nonlinear too, for example in stick-slip friction.

Actually, while MD simulations are %proving to be a
quite valuable %tool
in qualitatively catching the
physics of microscopic friction between extended solids, a %full
quantitative agreement with experimental
results is still beyond hopes %remains elusive
~\cite{VanossiRMP}.
Besides the practical difficulty posed by the necessity to describe inter-atomic interactions by either empirical force fields
or  with costly first principles calculations, an additional weak point of MD simulations lies %certainly
in the impossibility to access
the experimental time scales~\cite{mancolicani}. When attempting to simulate, e.g., a nanoscale Friction Force Microscopy experiment, with
the tip advancing at a far low average speed of $\simeq 1~\mu{\rm m/s}$, one can typically
simulate a miserable $\simeq 1$~pm advancement in a standard run, far too short to observe even a single
atomic-scale event, let alone reaching a steady state, or the development of any instability process,
and thus the quantitative evaluation of any useful frictional property.
Therefore, whenever long-distance correlations and/or slow diffusive phenomena
and/or long equilibration times are to be expected, fully atomistic MD approaches will only grab a qualitative scenario
of the system tribological response. Nevertheless, there is so much direct physical insight to be extracted from MD simulations
that it does make sense to run them even at larger speeds than in Atomic Force Microscopy (AFM) or Surface Force Apparatus experiments; and in fact, the sliding speed adopted
in most current atomistic MD frictional simulations is much higher, in the $0.1$ to $10$ m/s range.

The fast frictional motion in MD simulations ends up of course generating a vast amount of Joule heat. 
At the same time, the simulated system where that Joule heat is dispersed is generally of very limited size compared 
to the practically infinite environment of real experiments. That raises the problem, which is the focus of the present paper,
%for the partial lack of quantitative predictivity of MD techniques has to deal with the
of how that Joule energy can be continuously dissipated, ``thermostated'' away,  in order for the simulated system to reach a realistic steady
state rather than building up. %Until MD is used to study systems a
At equilibrium,
it does not matter how the thermostat scheme is built, because equilibrium properties do not depend on it.
On the contrary, in %studying
dynamical non-equilibrium processes, such as those occurring in tribology under the action
of external drive, the choice of a suitable physical thermostat is crucial, to dispose of the external energy
which is continuously pumped into the system. In the framework of wearless friction, for instance, sliding-induced
creation of phonons is a crucial mechanism of energy dissipation. An unsolved problem in realistic MD is that
the generated phonons cannot escape the small simulated contacting region between a slider and the underneath substrate
(see Fig.~\ref{figura1}) unlike in the real system, where they can properly disperse the Joule heat away from the interface.
The simulation cell boundaries back-reflect the phonons towards, e.g., the slider-substrate contact, as shown in panel (b)
of Fig.~\ref{figura2}, affecting so the frictional response. As phonons are continuously generated by sliding,
the simulated portions of the slider and substrate heat up, reaching quickly the melting point.
Thus, in order to attain a frictional steady state in simulation, the Joule heat must be removed. Unfortunately, a realistic
energy dissipation is generally impossible to mimic reliably, owing to size limitations of the simulation cell.
The empirical introduction in the equations of motion of ad-hoc Langevin viscous damping terms $-m \gamma \dot{q}_i$ (with
$m$ and $\dot{q}_i$ the mass and the velocity of the $i$-th substrate particle) and of an associated random noise, corresponding
to some ``thermostat'' temperature $T$~\cite{zwanzig}, represents the handiest and commonest solution,
which most simulations adopt. However,
both this procedure and the choice of thermostat and damping parameters $\gamma$ are vastly arbitrary.
The problem is not just one of principle, for in many cases (including, just as a significant example, multiple-slips
in AFM~\cite{carpick}) the resulting steady state and friction coefficient actually depend upon the choice of these unphysical parameters.
Here, after demonstrating  this unphysical dependence, we will pursue and detail a viable solution, whose core
was already outlined in a recent paper~\cite{BenassiPRB}.
\section{Non-Markovian Langevin approach for realistic tribological modeling}
Basically, one wishes to modify the equations of motion inside a relatively small simulation
cell so that they reproduce the frictional dynamics of a much larger system, once the remaining
variables are integrated out. Integrating out degrees of freedom is a traditional problem,
largely analyzed in the literature~\cite{zwanzig,maradudin,rubin1}.
In the context of MD simulation, Green's function methods were formulated for quasi-static mechanical contacts~\cite{muser};
approaches based on a discrete-continuum matching have also been discussed~\cite{luan2}. Among others,
time honored dissipation methods have been considered which replace the dynamics of the surrounding degrees of freedom
(the ``heat-bath'') by several terms in the equations of motion for the system, describing effects~\cite{zwanzig64,mori65,ReimannEvstigneev04}
such as 1) the renormalization of the forces acting on and between the relevant coordinates; 2) the introduction of viscous drag
describing the energy dissipation from the system into the heat bath; 3) the introduction of random forces describing
the inverse effect of energy transfer from the bath into the system. Recently~\cite{BenassiPRB} a direct implementation of a
non-Markovian dissipation scheme, based on early formulations by Magalinskii and Rubin~\cite{Magalinskii,rubin1} and subsequent
derivations by Li et al.~\cite{Li} and by Kantorovich~\cite{kantorovich1,kantorovich2}, has
demonstrated the correct disposal of friction-generated phonons
%
%...............................................................................................
\begin{figure}
\centering
\includegraphics[width=7cm,angle=0]{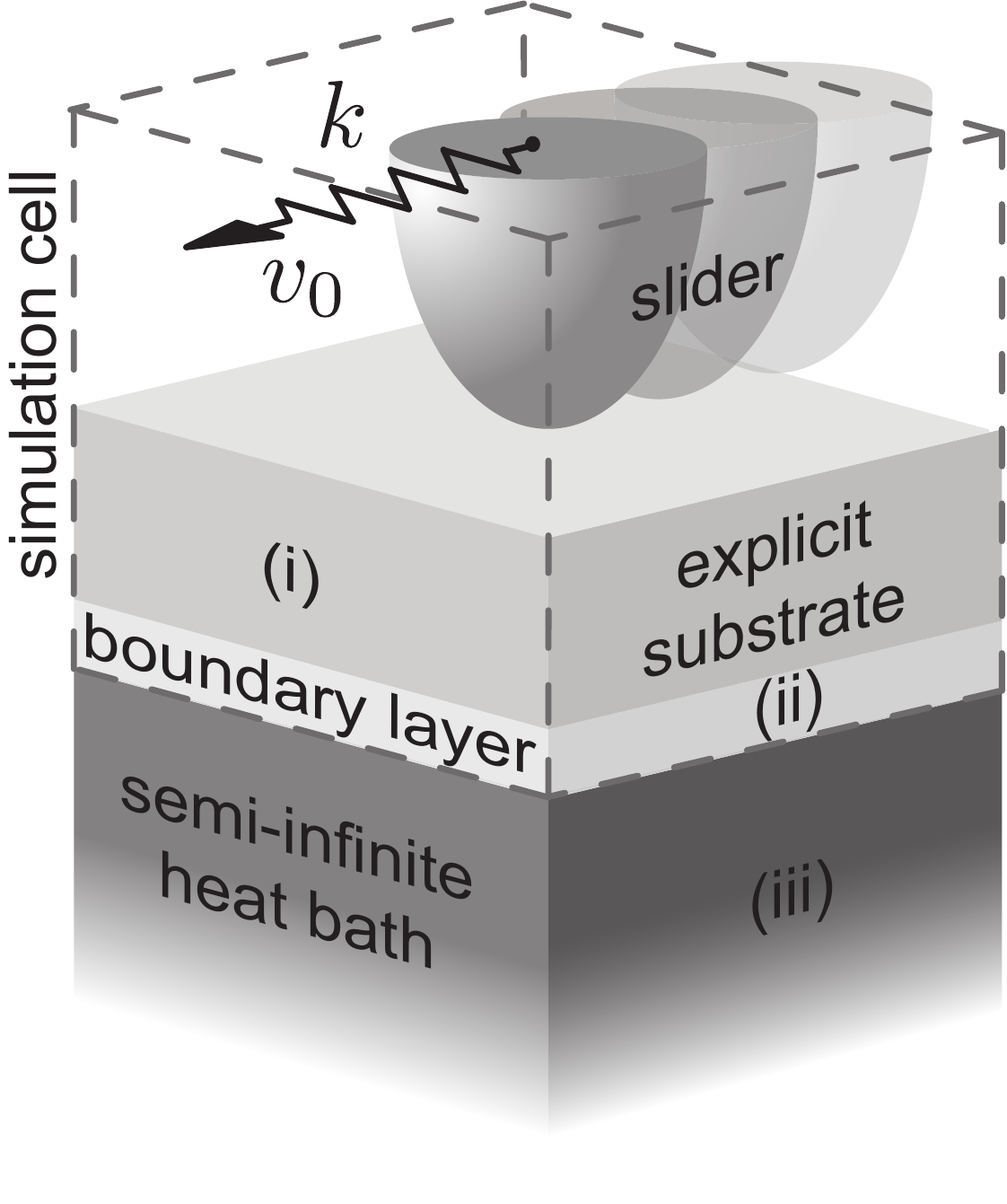}
\caption{%Schematic view of our non-markovian approach to the simulation of a typical sliding experiment.
Ideal block-scheme of a MD simulation of friction. To account properly for heat dissipation,
the infinitely-thick substrate is divided into three regions: (i) a `'live`' slab comprising layers whose atomic motion is fully
simulated by Newton's equations; (ii) a dissipative boundary zone, coincident with the deepmost simulated layer, whose
dynamics includes effective damping (e.g., non-Markovian Langevin-type) terms, as in Eq. (\ref{quasi}); (iii) the remaining semi-infinite
solid, acting as a heat bath, whose degrees of freedom are integrated out.}
\label{figura1}
\end{figure}
%...............................................................................................
%
%..........................................................................................................
\begin{figure}
\centering
\includegraphics[width=8.0cm,angle=0]{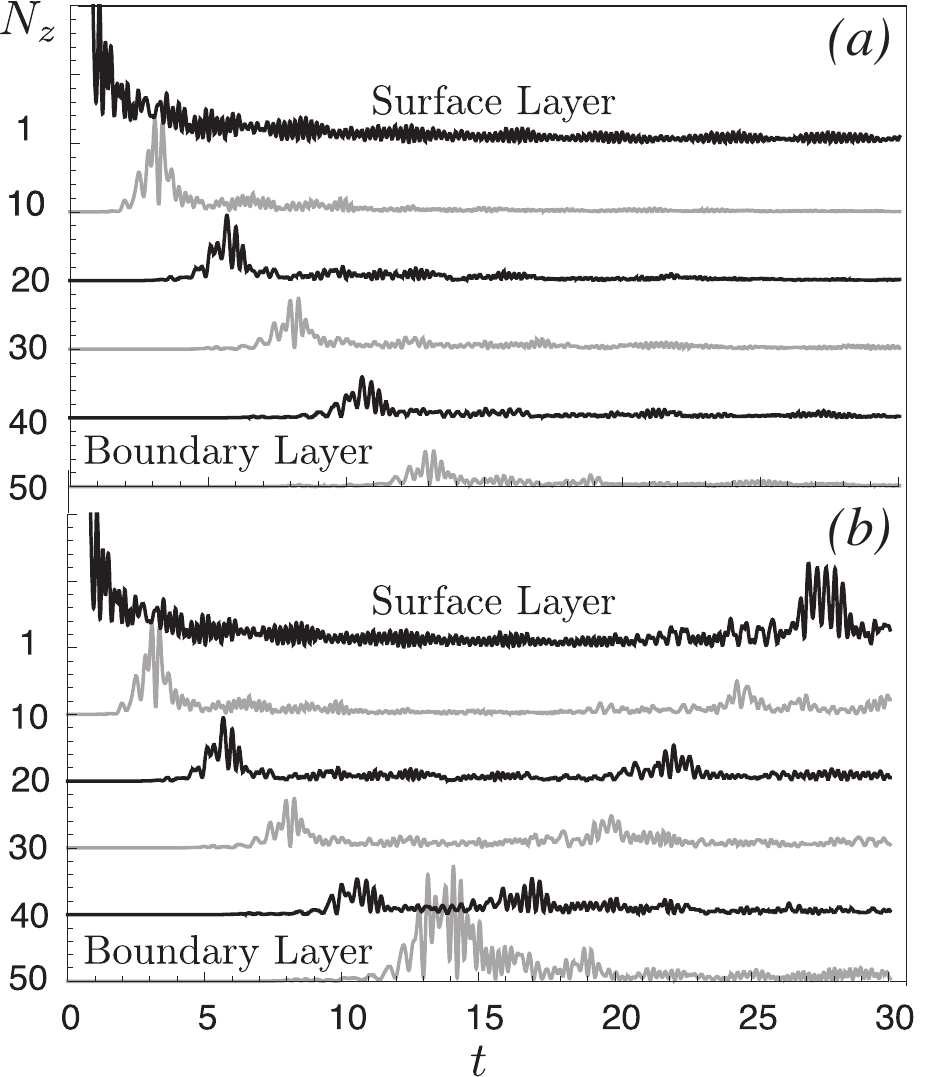}
\caption{Propagation into the substrate of surface injected energy. Tapping on the surface layer ($N_z=1$)
a burst of phonons has been created, its time evolution is monitored plotting the average kinetic energy of
equi-spaced atomic layers versus time. (a) shows a complete absorption of the phonon batch as it reaches the
bottom of the simulation cell ($N_z=50$) where our dissipation scheme is applied. (b) shows a total
back-reflection of phonons when the correct dissipative kernels are switched off.}
\label{figura2}
\end{figure}
%..........................................................................................................
in realistic MD simulations of sliding tribological systems,
as the one sketched in  Fig.~\ref{figura1}.
Once that was done, one could benchmark some simpler empirical Langevin scheme optimizing the $\gamma$ parameters
so as to yield less arbitrary frictional properties. We describe here in detail how both goals are achieved, picking for our
demonstration, without loss of generality, a specific two-dimensional (2D) realization.

We consider a simplified tribological system red where the upper slider is represented by a one-dimensional (1D) 
chain of atoms along the $x$-axis driven on top of a 2D semi-infinite crystalline substrate lying in the $(x,z)$ plane, where atoms
interact, for simplicity, via first-neighbor Lennard-Jones (LJ) potential. The slider, pressed against the substrate by a normal ``load''
$F_{0}$, is driven along $x$ (parallel to the surface) through a spring $k$, whose end is pulled at constant velocity $v_0$.
Following earlier formulations~\cite{kantorovich1}, the ideal infinitely thick substrate is divided, as sketched in in Fig.~\ref{figura1} 
in a 3D cartoon,
into three regions: (i) an explicitly simulated substrate portion of $N_z$ atomic layers with displacement vectors $\mathbf{r}(t)$,
(ii) the dissipative boundary layer, with displacement vectors $\mathbf{q}(t)$;
and (iii) the remaining semi-infinite solid acting as a phonon absorber, heat bath, with displacement vectors $\mathbf{b}(t)$.
Under certain, not too restrictive, assumptions described below, the heat bath degrees of freedom (iii) can be integrated out
to let a small simulation cell, namely (i)+(ii), account exactly for the energy dissipation as due to a semi-infinite substrate,
where the boundary layer (ii) is now ruled by effective non-Markovian Langevin equations, as derived in the following.
The first needed assumption is to substitute the full LJ potential within regions (ii) and (iii), i.e. far away from the sliding interface,
with its harmonic approximation.
This choice, necessary to derive an exact analytical form for the effective forces acting on the boundary atoms,
is all the more accurate the weaker the intensity of the slider perturbation and the lower the temperature.
Nevertheless, for crystalline substrates well below the Debye temperature, anharmonic perturbations reaching the heat bath 
can always be avoided by a sufficient thickness $N_z$ of the explicitly simulated substrate (i): these excitations, 
traveling through the LJ substrate, will gradually lose their energy turning into harmonic phonons prior approaching 
the boundary harmonic absorber.
In a compact matrix notation, the hamiltonian of the system reads
\begin{equation}
\mathcal{H}=\mathcal{T}+\mathcal{U}(\mathbf{r},\mathbf{q})+\mathbf{q}^{\dagger}\cdot \hat{\boldsymbol{\theta}}\cdot\mathbf{q}+
\mathbf{q}^{\dagger}\cdot \hat{\boldsymbol{\phi}}\cdot\mathbf{b}+\mathbf{b}^{\dagger}\cdot \hat{\mathbf{D}}\cdot\mathbf{b},
\label{hamiltonian}
\end{equation}
where $\mathcal{T}$ is the overall kinetic energy term, $\mathcal{U}$ is the LJ interactions among atoms in region (i)
and between regions (i) and (ii), $ \hat{\boldsymbol{\theta}}$ and $\hat{\boldsymbol{\phi}}$ are the LJ harmonic approximations
for the atomic interactions in region (ii) and between regions (ii) and (iii) respectively, and $\hat{\mathbf{D}}$
is the dynamical tensor of the heat bath (iii). Matrices and vectors have the form
\begin{equation}
\hat{\mathbf{D}}=\left( \begin{array}{cc}
\hat{\boldsymbol{D}}_{xx}& \hat{\boldsymbol{D}}_{xz}   \\
\hat{\boldsymbol{D}}_{zx} & \hat{\boldsymbol{D}}_{zz}
\end{array} \right),
\qquad
\mathbf{q}=\left( \begin{array}{c}
\mathbf{q}_{x}\\
\mathbf{q}_{z}
\end{array} \right),
\end{equation}
where each component is again a matrix or a vector of components $D_{\mu\nu}^{ij}$ or $q_{\mu}^{i}$, with latin indexes running
over the atoms and greek indexes running over the two $x$ and $z$ coordinates.
From the Hamiltonian (\ref{hamiltonian}), we can derive the following three sets of equations of motion:
\begin{align}
&m \ddot{\mathbf{r}}(t)=-\frac{d\,\mathcal{U}(\mathbf{r},\mathbf{q})}{d \mathbf{r}},\\
\label{seghetta}
&m \ddot{\mathbf{q}}(t)=-\frac{d\,\mathcal{U}(\mathbf{r},\mathbf{q})}{d \mathbf{q}}-\hat{\boldsymbol{\theta}}\cdot\mathbf{q}(t)-\hat{\boldsymbol{\phi}}\cdot\mathbf{b}(t),\\
&m \ddot{\mathbf{b}}(t)=-\hat{\boldsymbol{\phi}}\cdot\mathbf{q}(t)-\hat{\mathbf{D}}\cdot\mathbf{b}(t).
\label{sega}
\end{align}
Notice that the dynamics of atoms in region (i) is influenced only by atoms of region (ii), while the dynamics of region (iii)
depends only upon the dynamics of region (ii), in other words, thanks to the adopted cut-off LJ interaction, regions (i) and (iii)
are decoupled and they interact only indirectly via the boundary layer (ii).
The thickness size of the boundary layer (ii) depends on the cut-off radius: by considering only nearest-neighbors in the LJ interaction,
we end up in our case with a region (ii) made of a single atomic layer\cite{kantorovich2}.
Thanks to the assumed harmonicity of the heat bath interactions, we can decouple the equations for $\mathbf{b}(t)$,
diagonalizing the dynamical tensor $\hat{\mathbf{D}}$ and finding its eigenvalues $\omega_i$ and eigenvectors $\boldsymbol{\lambda}_i$.
By using the eigenvectors as a basis set $\mathbf{b}(t)=\sum_i \xi_i(t) \boldsymbol{\lambda}_i$, we substitute this projection
into Eq.(\ref{sega}), obtaining a set of easily solvable decoupled equations for the normal coordinates $\xi_i$,
\begin{equation}
\ddot{\xi}_i(t)+\omega^2_i\xi_i(t)=-\boldsymbol{\lambda}_i^{\dagger} \cdot \hat{\boldsymbol{\phi}} \cdot \mathbf{q}(t),
\end{equation}
where the rhs is a time dependent scalar quantity.
The final expression for $\mathbf{b}(t)$ becomes
\begin{widetext}
\begin{equation}
\mathbf{b}(t)=\sum_i \boldsymbol{\lambda}_i^{\dagger}\cdot
\bigg(
\mathbf{b}(0) \cos(\omega_i t)+
\dot{\mathbf{b}}(0) \frac{\sin(\omega_i t)}{\omega_i}
-\hat{\boldsymbol{\phi}}\cdot\frac{\mathbf{q}(t)}{\omega_i^2}
+\hat{\boldsymbol{\phi}}\cdot\mathbf{q}(0)\frac{\cos(\omega_i t)}{\omega_i^2}
+\hat{\boldsymbol{\phi}}\cdot \int_{0}^{t}\dot{\mathbf{q}}(s) \frac{\cos(\omega_i(t-s))}{\omega_i^2} ds
\bigg)\boldsymbol{\lambda}_i,
\end{equation}
\end{widetext}
which depends on the initial conditions of atoms in region (iii) and on the actual position of atoms
in region (ii). By substituting this expression into Eq.(\ref{seghetta}), we get
\begin{equation}
m \mathbf{q}(t)=-\frac{d\,\mathcal{U}(\mathbf{r},\mathbf{q})}{d \mathbf{q}}+\big(\hat{\mathbf{K}}(0) - \hat{\boldsymbol{\theta}}\big)\cdot\mathbf{q}(t) \\
-m\int_{0}^{t}\hat{\mathbf{K}}(t-s)\dot{\mathbf{q}}(s)ds+\mathbf{F}(t),
\label{quasi}
\end{equation}
where $\hat{\mathbf{K}}(t)$ and $\mathbf{F}(t)$ are defined as follows
\begin{align}
\label{oca}
&\hat{\mathbf{K}}(t) = \sum_{i}
 \bigg[ \frac{(\boldsymbol{\lambda}_i^{\dagger}\cdot \hat{\boldsymbol{\phi}})
              (\hat{\boldsymbol{\phi}}\cdot \boldsymbol{\lambda}_{i})}{\omega^2_i} \bigg] \; \cos{(\omega_i t)}, \\
& \mathbf{F}(t)=-\sum_i(\hat{\boldsymbol{\phi}}\cdot\boldsymbol{\lambda}_i)\boldsymbol{\lambda}_i^{\dagger}\cdot\bigg(
 \mathbf{b}(0)\cos(\omega_i t)+\dot{\mathbf{b}}(0)\frac{\sin(\omega_i t)}{\omega_i}
\bigg).
\label{bbb}
\end{align}
Equation (\ref{quasi}) still depends on the initial conditions of the heat bath through $\mathbf{F}(t)$.
Because this region is in principle infinitely extended, we cannot specify the initial conditions
for the position and the velocity of all its atoms; however, we are allowed to perform an equilibrium canonical ensemble average
introducing a temperature $T$.
Using for the partition function the bath hamiltonian only, it is easy to prove that
\begin{align}
&\langle \xi_i (0) \xi_j (0)\rangle=\boldsymbol{\lambda}_i^{\dagger} \cdot\langle\mathbf{b}(0) \mathbf{b}(0)\rangle\cdot\boldsymbol{\lambda}_j=\frac{K_B T }{m \omega_i^2}\delta_{ij},\\
&\langle \dot{\xi_i} (0) \dot{\xi_j} (0)\rangle=\boldsymbol{\lambda}_i^{\dagger} \cdot\langle\dot{\mathbf{b}}(0) \dot{\mathbf{b}}(0)\rangle\cdot\boldsymbol{\lambda}_j=\frac{K_B T}{m}\delta_{ij},\\
&\langle \xi_i (0) \dot{\xi_j} (0)\rangle=\boldsymbol{\lambda}_i^{\dagger} \cdot\langle\mathbf{b}(0) \dot{\mathbf{b}}(0)\rangle\cdot\boldsymbol{\lambda}_j=0,
\end{align}
being $K_B$ the Boltzmann's constant. Another possibility, adopted for example in ref.\cite{kantorovich1,kantorovich2}, is to include in the partition function also the term ruling the interaction between region (ii) and (iii). As a result the final effective equation of motion (\ref{meco}) takes a slightly different form.
Using the previous conditions into Eq.(\ref{bbb}), we end up with the following statistical properties
for the force $\mathbf{F}(t)$
\begin{equation}
\langle \mathbf{F}(t)\rangle=0,\qquad
\langle \mathbf{F}(t)\mathbf{F}(t')\rangle=m K_B T\; \hat{\mathbf{K}}(t-t'),
\end{equation}
or in component notation
\begin{equation}
\langle F^{i}_{\mu}(t) \rangle=0,\qquad
\langle F^{i}_{\mu}(t) F^{j}_{\nu}(t')\rangle = m k_B T K^{ij}_{\mu\nu}(t-t').
\label{regole}
\end{equation}
Thus Eq.(\ref{quasi}) can be regarded as a non-Markovian Langevin equation with a gaussian random noise
correlated according to the rules (\ref{regole}), and a dissipative term with a memory kernel function specified by (\ref{oca}).
Its expression in single component notation is given by
\begin{eqnarray} \label{meco}
m\ddot{q}^{i}_{\mu}(t) &=& -\frac{d\,\mathcal{U}(\mathbf{r},\mathbf{q})}{d\,q^{i}_{\mu}} -
  m \sum_{j,\nu} \int_{0}^{t} \! ds \; K^{ij}_{\mu\nu}(t-s) \;\dot{q}^{j}_{\nu}(s) \nonumber \\
   && + F^{i}_{\mu}(t) + \sum_{j,\nu} q^{j}_{\nu}(t) \bigg( K^{ij}_{\mu\nu}(0) - \theta^{ij}_{\mu\nu} \bigg).
\end{eqnarray}
The first term takes into account the interaction between the boundary layer atoms and the rest of the simulated substrate.
The second one is non-Markovian and non-conservative, introducing an effective damping proportional
to the velocity of all the boundary layer atoms, via a time convolution with the memory kernel functions $K^{ij}_{\mu\nu}(t)$.
The third term of Eq.(\ref{meco}) is the gaussian correlated noise ruled by the same memory kernel functions
involved in the dissipation, in agreement with the fluctuation-dissipation theorem.
Notice that in a standard Langevin equation the compliance with the fluctuation-dissipation theorem is imposed a priori
and the noise properties are derived from this constraint.
%..........................................................................................................
\begin{figure}
\centering
\includegraphics[width=8.5cm,angle=0]{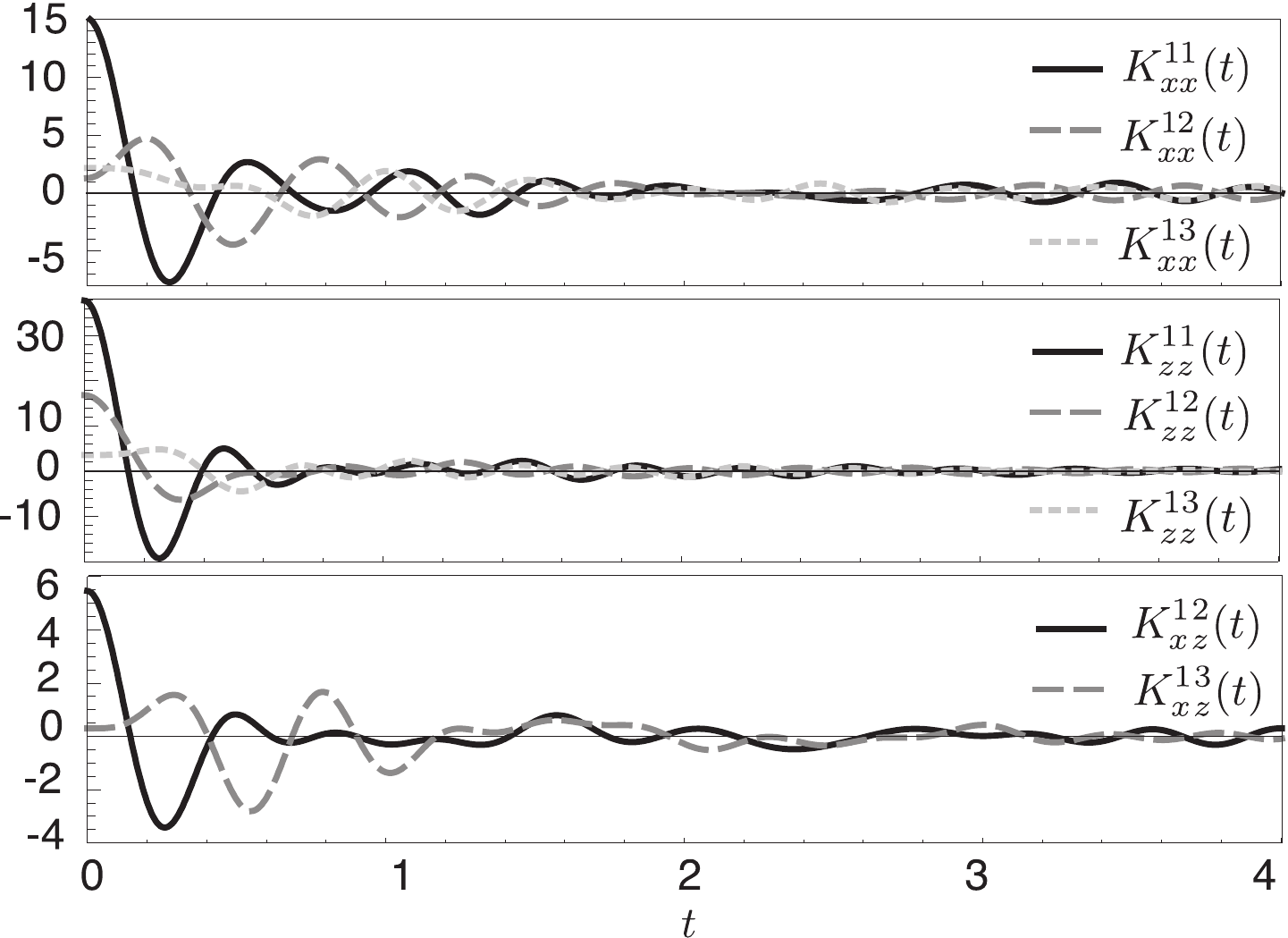}
\caption{Plot of some selected memory kernel functions versus time (LJ units).}
\label{figura3}
\end{figure}
%..........................................................................................................
%..........................................................................................................
\begin{figure*}
\centering
\includegraphics[width=15.0cm,angle=0]{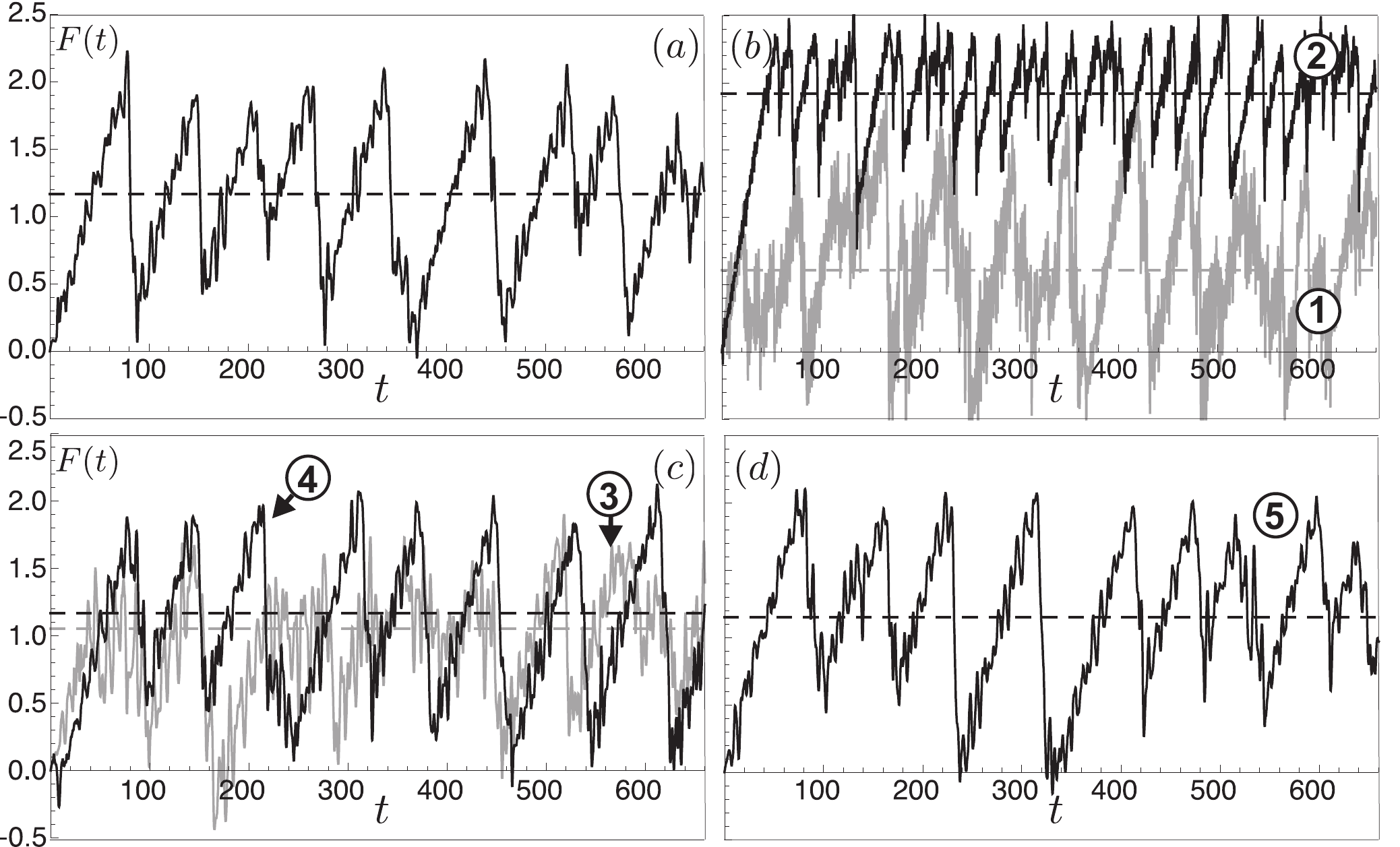}
\caption{(a) Calculated friction force profile $F(t)$ for the full non-Markovian dissipation scheme of Eq.~(\ref{meco}), and for different empirical viscous damping schemes (b),(c) and (d) described in text, and identified by numbers $1$-$5$ in Fig.\ref{figura5}. Dashed lines: mean value $\langle F \rangle$.}
\label{figura4}
\end{figure*}
%..........................................................................................................
In our formulation, which starts from a microscopic set of Hamilton's equations, the fluctuation-dissipation theorem
is automatically fulfilled just performing the canonical ensemble average.
The last term in Eq.~(\ref{meco}) finally represents the harmonic coupling between $i$-th and $j$-th atoms within the boundary layer,
where the coupling constant $\theta_{\mu\nu}^{ij}$ is modified by $K^{ij}_{\mu\nu}(0)$.
This renormalization of the elastic coupling for the region (ii) atoms vanishes as we include the interaction between the bath and the boundary layer 
into the partion function for the ensemble average.
It has been demonstrated theoretically \cite{kantorovich1}, and it can be easily verified in simulations,
that the application of Eq.~(\ref{meco}) to the boundary layer alone is sufficient to force the whole system to follow a canonical ensemble
distribution with temperature $T$.
The memory kernel matrix (\ref{oca}) in the single component notation reads
\begin{equation} \label{meco2}
K^{ij}_{\mu\nu}(t) = \sum_{k,l,m,\alpha,\beta}
 \bigg[ \frac{\big((\boldsymbol{\lambda}_k^{\dagger}) ^{l}_{\alpha}\; \phi_{\mu\alpha}^{il}\big)
              \big(\phi_{\nu\beta}^{jm}\; (\boldsymbol{\lambda}_{k})^{m}_{\beta}\big)}{\omega^2_k} \bigg] \; \cos{(\omega_k t)}.
\end{equation}
Each component is built from the harmonic eigenvalues and eigenvectors of the heat-bath dynamical matrix
and from the coupling vectors $\boldsymbol{\phi}_{\mu\alpha}^{il}$ containing the harmonic coupling constants
of the $i$-th atom of region (ii) with the $l$-th heat-bath atom.
As shown in Fig.~\ref{figura3}, the kernels oscillate and decay rapidly
with time, with power law tails due to the bath acoustical phonon branches.
However as long as the heat bath region remains finite the summations in (\ref{meco2}) are limited and the kernels are quasi-periodic functions\cite{zwanzig}. 
Waiting for a large time $\Lambda$, which depends on the heat bath size, the kernel functions rise and decay again repeatedly, 
this time periodicity marks the energy back-reflections from one end of the finite heat bath to the opposite one. 
In the limit of infinitely extended heat bath $\Lambda\rightarrow\infty$, no energy back reflection occurs. 
The numerical calculation of $\omega_k$ and $\boldsymbol{\lambda}_k$ can be carried out only for a finite dynamical tensor, 
i.e. for a finite bath, however we can set $K^{ij}_{\mu\nu} (t)=0$ for all $t>\tau$ with $\tau<\Lambda$ preventing the first reflection. 
If the heat bath is large enough we verified that $K^{ij}_{\mu\nu}(t)$ for $t<\tau$ is well converged, its shape
being insensitive to the addition of more terms in the summations of (\ref{meco2}).
By cutting kernels off after a time $\tau$ one can limit the time-integrals in Eq.~(\ref{meco}), which need to be evaluated at each time step, 
thus decreasing the heavy computational cost. But $\tau$ represents also the maximum time for which the boundary layer retains memory and correlation, 
therefore, via some convergence tests, we have to be sure that the quantities of interest do not depend on the chosen $\tau$ value.
Periodic boundary conditions along the $x$ direction guarantee translational invariance, so
that $K^{ij}_{\mu\nu}(t)$ is a function of $\vert i-j \vert$ only.
As kernels inherit their symmetry properties from those of the heat-bath dynamical matrix, one can
also show that $K^{ij}_{\mu\nu}(t)=K^{ij}_{\nu\mu}(t)$ and $K^{ij}_{\mu\nu}(t)=K^{ji}_{\nu\mu}(t)$.
When the separation $\vert i-j \vert$ grows, $\vert K^{ij}_{\mu\nu}(t)\vert$ decrease,
but again not exponentially, and correlations must be included up to large distance.
Implementing this set of equations, along with ordinary Newton's equations governing the remaining slider and
substrate atom motion was our first MD simulation step. Figure~\ref{figura2}(a) illustrates how a relatively thin
(i.e. $N_z=30$ layers) substrate (i+ii) is able to mimic the full ideal semi-infinite system (i+ii+iii).
Layer-resolved kinetic energies inside the simulated substrate show a group of phonons initially created
at the upper interface and propagating away from it. Upon reaching the boundary layer the phonons are perfectly absorbed
as they propagate into the (integrated-out) semi-infinite crystal (iii).
For comparison, Fig.~\ref{figura2}(b) shows the same phonons massively back reflected once the memory kernels
are removed from the boundary layer.
\section{Simulating atomic stick-slip}
We next proceed to simulate sliding friction by driving the slider (consisting, in the adopted 2D modeling,
of a LJ chain of $N'_x=9$ atoms) over the live substrate, consisting of $N_z=30$ close packed layers and $N_x=10$ atoms per layer.
Simulations were performed at temperature $k_BT=0.035$,
roughly corresponding to $T/T_{melting}=0.06$ (LJ units used throughout).
To favor sliding, the strength of the slider-substrate LJ interaction
is reduced from $1$ to $0.6$. The equations of motion are integrated by a modified velocity-Verlet
algorithm with a time step of $\Delta t=5\cdot 10^{-3}$, and the memory kernel functions are cutoff
at $\tau=5\cdot 10^{3}$ time-steps.
\footnote{
The correlated random noise sequence, to be applied to the boundary layer atoms, has been generated at the beginning of the simulation using the rules (\ref{regole}).
If we have to correlate in time a single random number sequence, we can generate a set of uncorrelated numbers in Fourier space, multiply them by the Fourier transform of the correlation matrix and make the inverse transform to get back to the real space\cite{noise}.}
Both the vertical load $F_0$ and the lateral driving are applied to the slider center of mass, the equation of motion for the slider degrees of freedom $\mathbf{s}_i$ is
\begin{align}
&m \ddot{s_{ix}}=-\frac{d\,\mathcal{U}(\mathbf{r},\mathbf{s})}{d\,s_{ix}} -\frac{d\,\mathcal{U}(\mathbf{s})}{d\,s_{ix}}-k(s_{xCM}-v_0 t)\\
&m \ddot{s_{iz}}=-\frac{d\,\mathcal{U}(\mathbf{r},\mathbf{s})}{d\,s_{iz}} -\frac{d\,\mathcal{U}(\mathbf{s})}{d\,s_{iz}}-F_0
\end{align}
where $\mathcal{U}(\mathbf{r},\mathbf{s})$ is the LJ interaction with the substrate atoms (i) and $\mathcal{U}(\mathbf{s})$ is the LJ interaction among the slider atoms, $s_{xCM}$ is the slider center of mass position along $x$. As usual the friction force is measured by the spring elongation $k(s_{xCM}-v_0 t)$ representing the slider resistence to the lateral driving.
The applied load is $F_{0}=10$, the average sliding velocity $v_0=0.01$, and the spring constant $k=5$.
The result is the sawtooth force profile in Fig.~\ref{figura4}(a) typical of intermittent stick-slip friction.
The friction coefficient, obtained by averaging over several stick-slip events, is $\langle F \rangle/F_{0} = 0.116 \pm 0.002$.
The slider is slightly incommensurate with the substrate, so that the sawtooth pattern is quite irregular with a periodicity
not exactly matching one lattice spacing.
An anti-kink (physically corresponding to a tiny localized expansion in the particle array density of the slider 
due to the interface mismatch) appears at the interface, moving in the opposite direction with respect
to the slider: the height of the sawtooth spikes is proportional to the jump length of the anti-kink.
Higher spikes occur for simultaneous forward jumps of many atoms, smaller ones correspond to jumps of $2-3$ atoms at once.
A measure of the distribution of the spike heights is the variance of $F(t)$, i.e.,
\begin{equation}
\sigma = \frac{1}{\tau_s}\int_0^{\tau_s} [F(t)-\langle F \rangle]^2 dt,
\label{variance}
\end{equation}
where $\tau_s$ is the total simulation time. Numerical simulations carried out with the full Eq.~(\ref{meco}),
and the corresponding frictional results are essentially exact for the system considered.
That completes our first important goal of implementing the correct Joule heat removal, thus also establishing a benchmark reference.
Not surprisingly, this numerical implementation is time consuming. In particular the computational effort
required to integrate the non-Markovian term, where boundary atoms are strongly correlated,
scales as $N_x^2$. Carrying out future fully realistic frictional simulations for large-size 3D sliding systems
within this scheme is in our view entirely possible, but may pose some practical challenge of parallel computing.

%..........................................................................................................
\begin{figure}
\centering
\includegraphics[width=8.5cm,angle=0]{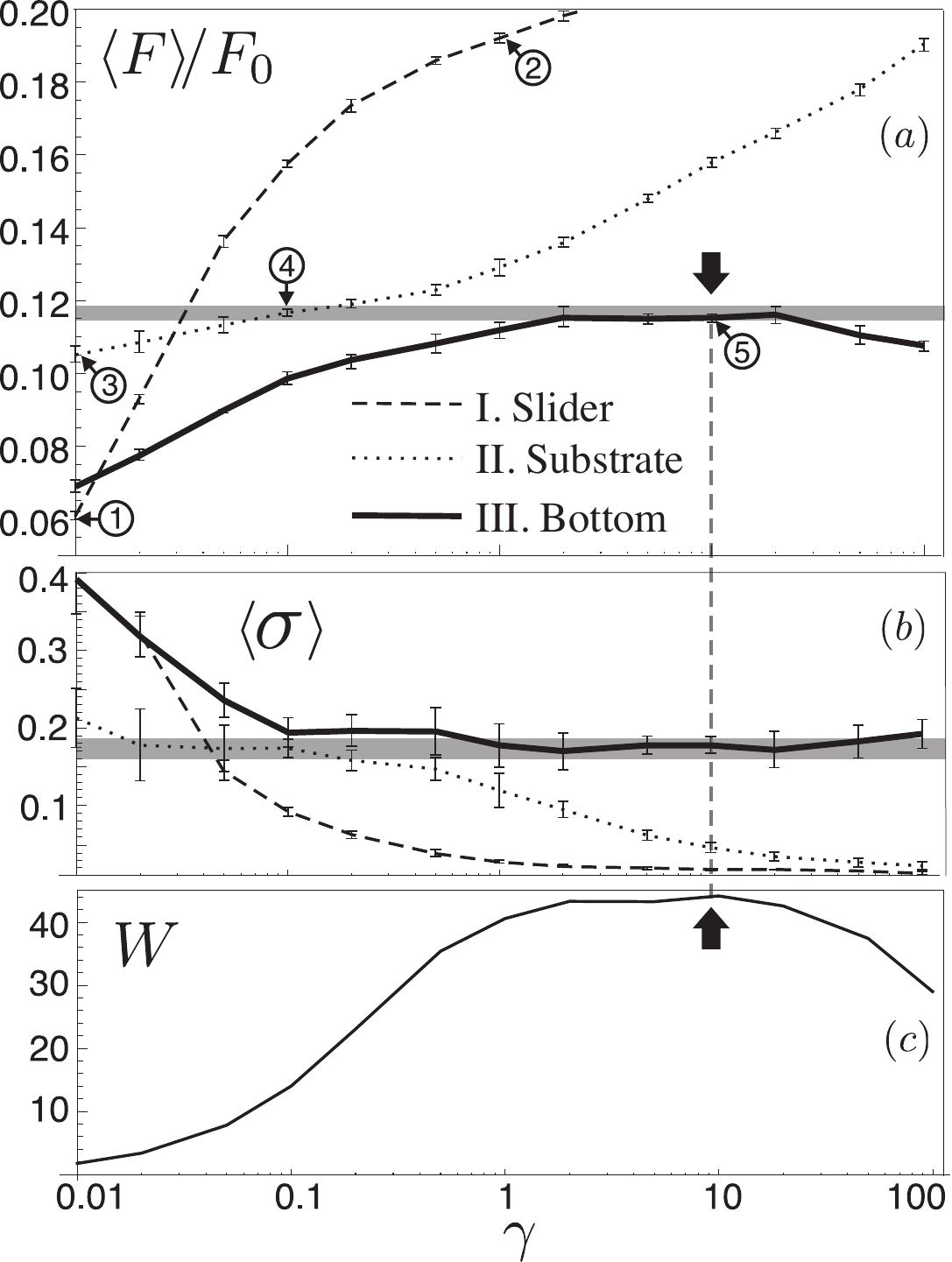}
\caption{(a) and (b) illustrate the friction coefficient $\langle F \rangle/F_0$ and variance $\langle \sigma \rangle$ behaviors as a function of
the damping coefficient $\gamma$ for different empirical Langevin dissipation schemes, in comparison with the exact values from
the full non-Markovian simulation (gray stripes). (c) shows the boundary layer absorbed energy $W$ of the Langevin thermostat (\ref{semplice}). Note the
good coincidence of exact and empirical frictional behavior for the optimal $\gamma$ that maximizes $W$.}
\label{figura5}
\end{figure}
\begin{figure}
\centering
\includegraphics[width=8.5cm,angle=0]{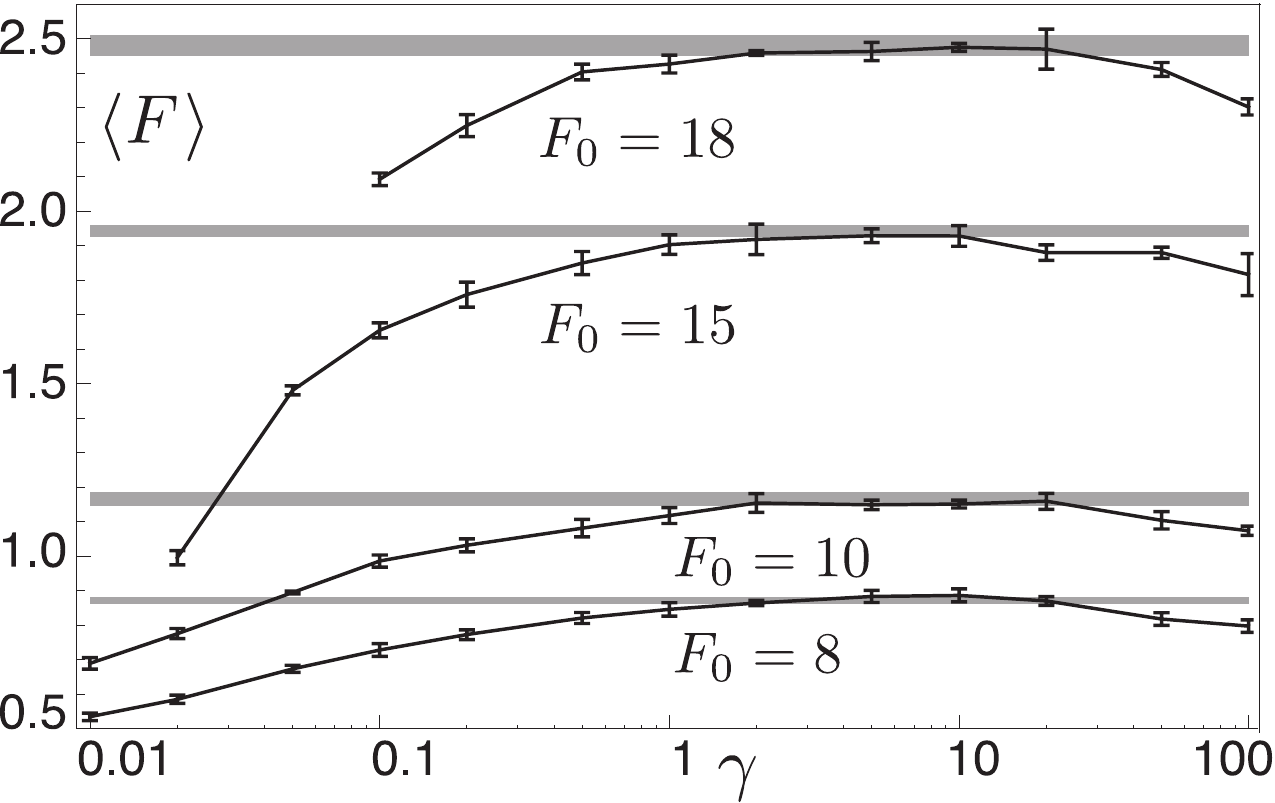}
\caption{Average friction force for different
loads $F_0$. Gray stripes show the values obtained with the non-Markovian approach in comparison with the black curves for the boundary Langevin scheme at different $\gamma$.}
\label{figura6}
\end{figure}

%..........................................................................................................
This brings us to our second point. As was mentioned, much simpler and faster approximate frictional simulations are
realized once the non-Markovian memory kernels of Eq.~(\ref{meco}) are empirically replaced with a more ordinary
Markovian Langevin viscous damping $-m\gamma \dot{q}^i_{\mu}(t)$, along with the appropriate gaussian stochastic force
$R_i(t)$ with $\langle R^i_{\mu}(t) \rangle = 0$ and $\langle R^i_{\mu}(t)R^{j}_{\nu}(t') \rangle = 2 m k_B T \gamma\delta_{\mu,\nu}\delta_{i,j}\delta(t-t')$,
so that the equation of motion of the $i$-th thermostated atom in the system reads
\begin{equation}
m \ddot{q}^{i}_{\mu}(t)=-\frac{d\,\mathcal{U}}{d q^i_{\mu}} - m \gamma \dot{q}^i_{\mu}(t) + R^i_{\mu}(t) \;,
\label{semplice}
\end{equation}
where $\mathcal{U}$ is the LJ inter-atomic interaction.
Performing a series of simulations with the same system parameters, the previous exact implementation
now offers the possibility to benchmark the empirical damping $\gamma$.
In principle, this standard Langevin scheme can be differently exploited, applying it to:
I. (Fig.\ref{figura4}(b), curves $1$ and $2$) the slider atoms only, while freezing the substrate degrees of freedom,
as typically done in the simplified framework of Prandtl-Tomlinson and Frenkel-Kontorova modeling~\cite{vanossi};
II.  (Fig.\ref{figura4}(c), curves $3$ and $4$)  to each substrate atom, possibly by making it site-dependent \cite{braunRep};
III.  (Fig.\ref{figura4}(d), curve $5$) just to the bottom simulation-cell layer, as considered for the parameter-free, non-Markovian, stochastic dynamics.
In all these cases, we find a strong dependence of the system frictional response on the choice of the damping $\gamma$,
in general deviating always systematically from the correct benchmark.
Figure~\ref{figura5}(a),(b) shows the behavior of the friction coefficient and its variance, respectively, as a
function of $\gamma$. The grey stripes indicates the benchmark values of $\langle F \rangle/F_0$ and $\langle \sigma \rangle$
obtained with our parameter-free dissipation scheme, mimicking a semi-infinite substrate.
The dashed line represents the results for standard Langevin equations applied, only, to the slider atoms (case I.):
this turns out to be the most unrealistic and $\gamma$-sensitive situation.
A too large $\gamma$ introduces a strong viscous character, and leads to overestimating the friction force,
while a too small $\gamma$ results in a chaotic behavior, with the slider dynamics being unable to dissipate enough energy.
This scenario is outlined in Fig.~\ref{figura4}(b) where the $F(t)$ stick-slip profile is plotted for $\gamma=0.01$
(blue solid line) and for $\gamma=1.0$ (black dotted line). At $\gamma=0.035$ in Fig.~\ref{figura5}(a), this average friction force curve
crosses the ``exact-method'' (grey) stripe with a value $\langle F \rangle/F_0=0.117\pm 0.001$,
but with a too large variance $\langle \sigma=0.28 \rangle$ (Fig.~\ref{figura5}(b)),
and a consequent very inaccurate reproduction of the stick-slip pattern (not shown).
The dotted line in Fig.~\ref{figura5} represents the Markovian Langevin thermostat applied, more realistically, to all substrate
atoms (case II.): the slider exchanges energy with the substrate by exciting phonons at the interface; these phonons are then damped
within the substrate independently of the slider velocity. However, a too large $\gamma$ will lead to a very viscous surface preventing
the correct energy exchange between the slider and the substrate, and $\langle F \rangle/F_0$ increases too much, as in the previous case.
A too small $\gamma$, on the contrary, makes the substrate unable to dissipate the phonons, which are then reflected back,
reaching again the surface and heating it to unphysically large temperatures, thus spuriously decreasing the friction force.
Fig.~\ref{figura4}(c), corresponding to such case II., shows $F(t)$ for a low $\gamma$ value of $0.01$ (blue solid line):
the effect of the reflected phonons is to reduce the static friction force, decreasing the swing of the saw-tooth profile.
$F(t)$ is also displayed for $\gamma=0.1$ (black dotted line): the average friction force here approaches
our semi-infinite substrate result, mimicking well also the stick-slip profile, as highlighted by the simultaneous good values of
the friction coefficient and the standard deviation in Fig.~\ref{figura5}(a),(b). However, there is here (case II.) no a priori
possibility to choose the optimal value of the damping parameter without having previously performed an exact non-Markovian benchmark
calculation.
Besides, in order not to directly interfere with the detailed dynamics and the slider-substrate energy exchange,
the Langevin viscous damping term should be switched on far from the surface as, e.g., in the bottom dissipation
layer (case III.), shown by the continuous line in Fig.~\ref{figura5}.
We find that there exists an optimal damping $\gamma_{opt}$ (here $\gamma_{opt} \sim 10$) for which both
the friction coefficient and its variance agree well with the exact values (see Fig.~\ref{figura5}(a),(b)).
Moreover, also the stick-slip profile in Fig.~\ref{figura4} for $\gamma=\gamma_{opt}$ (panel(d))
compares excellently with the exact one (panel(a)).
Remarkably the $\gamma$ value for which the friction profile better resemble the exact one corresponds to the one which maximizes the average friction force. 
In order to understand this relation, we look at the energy dissipated by the boundary layer:
\begin{equation}
W=-m \sum_{i}\int \gamma\; \dot{\mathbf{q}}^i\cdot d\mathbf{q}^i=-m \sum_{i}\int \gamma\; \vert\dot{\mathbf{q}}^i\vert^2 dt,
\end{equation}
finding a maximum at the same $\gamma$ values as illustrated in Fig.~\ref{figura5}(c).
This maximum occurs because back-reflection of phonons is large {\it both} when the boundary layer damping $\gamma$ 
is too small and too large.  The efficiency in the energy removal goes as $-m\gamma \dot{q}_i$, so that
at low $\gamma$ values the boundary layer atoms cannot dissipate significantly even if vibrating very fast; 
in the opposite limit of large $\gamma$, the boundary layer dynamics becomes so viscous (low atomic velocities) 
that an effective dissipation is again hampered.
At $\gamma=\gamma_{opt}$, we reach a good compromise between the strength of the damping and the atom velocities 
and most of the impinging energy is disposed of.
The agreement between the exact frictional results, where no phonons are back reflected, 
and the approximate ones is therefore best when energy back reflection is minimal 
and this can occur for a single $\gamma$ value only. 
The minimal phonon back reflection condition also establishes the smallest temperature at the sliding interface. 
%thus leading to a maximum in the friction force, atomic stick-slip friction is in fact known to decrese with temperature.
%QUI BISGNA CITARE QUALCUNO}
While this makes good physical sense, we still contemplate the possibility that the numerical
result might be just some kind of coincidence in a single simulation. 
We therefore proceed to change system parameters, including sliding velocity, and load.
In all cases we find an optimal $\gamma$ value, where both the friction force and the energy dissipated by the boundary layer are maximized and where both average friction and variance
coincide with the exact value separately calculated by a full non-Markovian simulation.
For example the variable load results of Fig.~\ref{figura6} show that the coincidence of optimal and exact friction is systematic as well as the presence of  the force maximum that can be thus exploited as a tool to calibrate the viscous coefficient $\gamma$ for any general system even without the exact non-Markovian benchmark.
%..........................................................................................................
%..........................................................................................................
\section{Conclusions}
We have shown here that sliding friction obtained by Molecular Dynamics simulations may depend heavily
on the scheme adopted for the elimination of Joule heat.
None of the empirical but commonly used  dissipation schemes seems satisfactory. One might for example
apply a Langevin viscous damping $\gamma$ to the slider atoms
alone~\cite{vanossi}, or, uniformly to all substrate atoms~\cite{luan}. Shown as dashed and dotted lines
respectively in Fig.~\ref{figura3},
the friction coefficients produced by these approximations, although crossing
the correct values as a function of $\gamma$, generally yield a much lower quality description
as seen by the stick-slip profiles in Fig.~\ref{figura2}.
More importantly, these schemes generally offer no clue on how to optimize the empirical parameter
$\gamma$ in the absence of the exact simulation.\\
We then showed how the real dissipation of
phonons into a harmonic semi-infinite solid substrate can be simulated by implementing well established
non-Markovian schemes.  Once the exact non-Markovian dissipation is replaced by an approximate and
empirical Langevin damping $\gamma$ applied to the bottom layer of the simulated substrate slab,
an optimal value for $\gamma$ is easily and variationally found by maximizing dissipation --
a condition which can be established without resort to any exact reference calculation.
This is a result which in all likelihood appears more general than the simple model used to demonstrate it,
and should thus be quite valuable for general applications.
\section{Acknowledgments}
A discussion with L. Kantorovich is gratefully acknowledged.
This work is part of Eurocores Projects FANAS/AFRI, sponsored by the Italian Research Council (CNR), and of FANAS/ACOF.
It is also sponsored by the Italian PRIN Contracts No. 20087NX9Y7 and No. 2008Y2P573, and by the Swiss National Science Foundation
SINERGIA Project CRSII2 136287$\backslash 1$.

\bibliographystyle{cj}
\bibliography{papero}
\end{document}